\documentclass{article} 
\newcommand{\re}[1]{(\ref{#1})}
\newcommand{\q}{\quad}

\newcommand{\pl}{\partial}
\newcommand{\pr}{^\prime}
\newcommand{\la}{\langle}
\newcommand{\ra}{\rangle}
\newcommand{\Ttilde}{\tilde{T}}
\newcommand{\Ktilde}{\tilde{K}}
\newcommand{\Etilde}{\tilde{E}}
\newcommand{\dStilde}{\tilde{dS}}
\newcommand{\atilde}{\tilde{a}}
\newcommand{\utilde}{\tilde{u}}
\newcommand{\tautilde}{\tilde{\tau}}
\newcommand{\taut}{\tilde{\tau}}

\begin{document}
\title{Energy Radiation of Charged Particles 
in Conformally Flat Spacetimes}
\author{Stephen Parrott
}
\maketitle
\begin{abstract}
\noindent
{\bf Original abstract:}
Consider the worldline of a charged particle in a static spacetime.
Contraction of the time-translation Killing
field with the retarded electromagnetic energy-momentum tensor gives a 
conserved electromagnetic energy vector which can be used to define 
the radiated electromagnetic energy.  This note points out that for 
a conformally flat spacetime, the radiated energy is the same as 
for a flat spacetime (i.e. Minkowski space).  
This appears to be inconsistent with an equation of motion for such particles 
derived by DeWitt and Brehme \cite{dewitt/brehme}
and later corrected by Hobbs \cite{hobbs}.
[End of original abstract]
\\[1.5ex] 
{\bf New abstract:}  Same as old abstract with last sentence deleted. 
The body of the paper is the same as previously.
A new Appendix 2 has been added discussing implications 
to the previous arguments  
of recent work of Sonego (J. Math. Phys. 40 (1999), 3381-3394)
and of Quinn and Wald (Phys. Rev. D 60 (1999), 
http://gr-qc/9610053).
\end{abstract} 
\section{Energy radiation in conformally flat spacetimes is the same
as in Minkowski space.}
Consider a spacetime which admits a Killing vector field $K = K^i$.
If $T = T^{ij}$ is any symmetric tensor
with vanishing covariant divergence, such as an energy-momentum tensor,
then the vector $E^j := K_\alpha T^{\alpha j}$ has vanishing divergence,
and so defines a ``conserved quantity'' (\cite{sachs/wu}, p. 96).
In Minkowski space one obtains conservation of energy-momentum in this
way from the Killing fields corresponding to space-time translations.

Now consider a static spacetime, which means
that there is a ``static'' coordinate system $x^i$ in which 
the metric coefficients
$g_{ij} = g_{ij} (x^1 , x^2 , x^3 )$ are independent of the ``time''
coordinate $x^0$ and that $g_{0I} = 0 $ for spatial indices $I = 1,2,3$.
\footnote{We follow the standard 
convention of using upper indices
for contravariant vectors and lower indices for covariant vectors, 
with repeated upper and lower indices assumed summed unless otherwise
indicated. 
Local coordinates are denoted by upper indices. The symbol ``:='' means
``equals by definition''.}
The infinitesimal generator $\partial_0$ of time translation
is then a Killing vector field, and this defines as above 
a conserved quantity $E^j := g_{00}T^{0j}$ which
is usually interpreted as electromagnetic ``energy'' when $T$ is the 
usual energy-momentum tensor of the electromagnetic field.  
There are special cases in which
this interpretation may not be appropriate (cf. \cite{parrott2}), but in
this paper, we'll follow tradition and 
call this conserved quantity ``energy''.
In this paper, ``radiation'' always means energy radiation unless otherwise
specified. 
We want to study the energy radiation of a charged particle with a 
prescribed worldline, such as a freely falling particle.

It is ``well known'' that a stationary charged particle 
in a static spacetime  does not radiate.%
\footnote{The phrase ``widely believed'' in place of ``well known''
would be more nearly accurate, but would also suggest 
a possibility of doubt which probably does not exist. 
That stationary particles do not radiate has never been rigorously proved
in reasonable generality, but seems to be universally believed. 
Further discussion and a proof under certain auxiliary 
hypotheses can be found in \cite{parrott2}, Appendix 2.}
A stationary particle is accelerated relative to local inertial frames,
but is {\em apparently} unaccelerated with respect to the
static coordinate frame.  On the other hand, a freely falling particle
is (by definition) unaccelerated with respect to local inertial frames
but {\em is}, in general, 
apparently accelerated with respect to the static coordinate 
frame.  One might guess that if a stationary (but accelerated) particle
does {\em not} radiate, then a freely falling particle should
radiate, the idea being that it may be acceleration relative 
to the static frame that determines radiation 
rather than acceleration relative to local inertial frames.  
This is not necessarily at variance with ideas of general covariance because 
singling out the time-like Killing vector field $\partial_0$ in 
a static spacetime destroys general covariance.

This problem has, in our opinion, never been satisfactorily treated in
the literature.  Unfortunately, this paper does not furnish a 
general analysis, but we do obtain an answer for any 
spacetime whose metric is of the special form: 
\begin{equation}
\label{eq1}
g_{ij}dx^i dx^j  = k(x^1 , x^2 , x^3 )^2 \left[ (dx^0)^2 - 
\sum_{J=1}^3 (dx^J )^2 \right]
\q ,
\end{equation}
that is, for any static, conformally flat metric.  
The essence of the answer 
is that a charged particle will radiate when and only when 
it is accelerated with respect to the ``static'' coordinate system
$x^i$ with respect to which the static metric \re{eq1} is written. 
In particular, a freely falling particle can radiate. 

For certain conformal factors $k$ in \re{eq1},
this is at variance with the main results
of \cite{dewitt/brehme}, \cite{hobbs}, \cite{hobbs2}. 
They derive an equation of motion
from considerations of energy-momentum balance, in effect calculating
energy-momentum radiation without explicitly calling it that. 
After correction by \cite{hobbs} of an error 
in \cite{dewitt/brehme}, 
this equation of motion (our equation \re{hobbseq} below)
looks like the Lorentz-Dirac equation with an additional
term involving the Ricci tensor.  
The method of \cite{hobbs2} leads to the conclusion that if a metric $g$
of form \re{eq1} is such that the this additional term vanishes, then 
a charged particle can follow a geodesic 
(i.e., the same ``freely falling'' 
motion that an uncharged particle would have).
This implies that there is no ``radiation reaction'' force and
presumably, no radiation.  

Thus the definition above of ``energy radiation'' is in this situation
in conflict with the definition implicit in \cite{dewitt/brehme}
and \cite{hobbs}.   
Both definitions are commonly used.
Neither definition is unreasonable, 
but neither is beyond question.  
We discuss possible objections to both definitions 
but advocate neither.
The purpose of this paper is simply to point out that 
the two definitions are sometimes (and probably usually) inconsistent. 
  
Let $F = F_{ij}$ denote the electromagnetic field tensor,
and $T = T^{ij} = {F^i}_{\alpha}F^{\alpha j} - (1/4) 
F^{\alpha \beta}F_{\alpha \beta} g^{ij}$ 
its corresponding energy-momentum tensor. 
The tensor $T$ implicitly depends
on the conformal factor $k^2$: denote by $\Ttilde$ the corresponding
tensor for $k \equiv 1$; i.e., for Minkowski space.  
More generally, when we compare various tensors in the spacetime
\re{eq1} with conformal factor $k^2$ with the corresponding tensors 
in Minkowski space, 
we'll consistently denote the Minkowski space 
versions by tildes as above.  
Index raising and lowering of tensors will be defined relative to
the relevant metric; for example, denoting the Minkowski metric 
by 
\begin{equation}
\label{minkowski}
(\tilde{g}_{ij}) := 
\mbox{diag}(1,-1,-1,-1) 
\q ,
\end{equation}
we define
${\Ttilde^i}_{~j} := \Ttilde^{i\alpha} 
\tilde{g}_{\alpha j}$.

We want to compare $T$ and $\Ttilde$.
Because of the conformal invariance of Maxwell's equations,
the 2-covariant form $F_{ij}$ of $F$ is independent of the
conformal factor $k^2$, but each index-raising with respect to $g$ 
introduces a factor
of $k^{-2}$ relative to the same index-raising in Minkowski space, 
and hence $T^{ij}$ differs from the corresponding tensor
$\Ttilde^{ij}$ for Minkowski space by a factor
$k^{-6}$:  $T^{ij} = k^{-6} \Ttilde^{ij} $.  

Let $K = K^i$ denote the Killing vector corresponding to
time translation; in differential-geometric notation,  $ K = 
\partial_0$.  For future reference, note that $K^i$ is independent
of the conformal factor $k^2$, but $K_i := g_{i \alpha} K^\alpha = k^2 
\Ktilde_i$. 
To any such Killing vector field $K$ is associated a ``conserved''
(i.e. zero-divergence) vector field $E^i := K_{\alpha}T^{\alpha i}$.
For our Killing field $K := \partial_0$, 
the integral of the normal component of 
$E$ over any spacelike submanifold is interpreted 
as the energy in the submanifold.
A similar integral over the three-dimensional timelike submanifold $S$ obtained by letting
a two-dimensional surface (such as a sphere) evolve through time
is interpreted as the energy radiated through the surface over the
time period in question.  Such integrals will be denoted
$\int_S E^\alpha \, dS_\alpha$.  
A fuller discussion is given in \cite{parrott2}, and 
precise mathematical definitions 
can be found in  \cite{parrott}, Section 2.8. 

The relation between the above integral in the spacetime
\re{eq1} and in Minkowski space is:
\begin{equation}
\label{eq2.5}
\int_S E^\alpha \, dS_\alpha = 
\int_S k^{-4} \Etilde^{\alpha} k^4 \dStilde_\alpha = 
\int_S \Etilde^{\alpha} \, \dStilde_\alpha 
\q .
\end{equation}
For a timelike manifold $S$ consisting of a spacelike surface evolving
through time, this says that
the energy radiation through the surface is {\em independent} of the conformal
factor $k^2$.  

Since the cancellation of the factors of $k$ in \re{eq2.5}
greatly simplifies our considerations, 
we would like to understand this with as little 
explicit calculation as possible.  
To understand the symbolic substitution $dS_\alpha = 
k^4 \dStilde_\alpha$, think of 
an integral like $\int_S E^\alpha \, dS_\alpha $ as obtained
by the following process. Imagine 
decomposing the three-dimensional submanifold $S$ into a large number
of 3-dimensional ``cubes'', 
each spanned by three tangent vectors to the submanifold. 
\footnote{Of course, tangent vectors don't actually ``lie in'' the submanifold,
but this is the intuitive picture on the infinitesimal level.  Also, we
use ``cube'' in place of the more cumbersome term ``parallelipiped''.
} 
From the corner of each 3-cube protrudes a vector $E$.  The three spanning
vectors for the cube together with $E$ span a 4-``cube''.  The integral is
thought of as the sum of the volumes of these 4-cubes, where ``volume''
is defined via the natural volume 4-form associated with the metric.
For a 4-cube spanned by four orthogonal vectors with nonzero norms, 
the volume is the
product of the four norms; hence the factor $k^4$ in 
$dS_\alpha = k^4 \dStilde_\alpha $.  The compensating factor $k^{-4}$ in
$\Etilde^i = k^{-4} E^i$ can be seen by looking at 
$E^i := T^{i\alpha} K_\alpha$.  As noted above, passing from $T$ to $\Ttilde$
involves three index-raisings of $F_{ij}$, 
which introduces a factor of $k^{-6}$, whereas
$K_\alpha = k^2 \Ktilde_\alpha$.  

In short, the energy radiation of a particle with an arbitrary worldline
is {\em the same as in Minkowski space}, 
which is well-known (\cite{parrott}, 
p. 160) to be essentially given by the integral of the square 
of the proper acceleration
over the worldline.  More precisely, if the particle has charge $q$
and is unaccelerated (in Minkowski space, not with respect to the metric
\re{eq1}) 
in the distant past and future, then for a particle with 
Minkowski space four-velocity $\utilde (\tautilde )$ at Minkowski
proper time $\tautilde$ and Minkowski  
proper acceleration 
$\atilde := d\utilde /d\tautilde$, the total energy 
radiation over all time
is:  
\begin{equation}
\label{eq3}
\int_S E^\alpha \, dS_\alpha = 
\int_S \Etilde^\alpha \, \dStilde_\alpha =  - \frac{2}{3}q^2 
\int_{-\infty}^{\infty}
\atilde^2
 \utilde^0 
\, d\tautilde 
\q ,
\end{equation}
where
$\atilde^2 := \atilde^\alpha 
\atilde_\alpha := \atilde^\alpha \tilde{g}_{\alpha\beta} \atilde^\beta$. 

The reason for assuming that the particle be unaccelerated in distant
past and future is that only under this hypothesis do all commonly
used calculational methods give identical results.  (Further explanation
will be given in Section 2, and a complete treatment can be found in
\cite{parrott}, Chapter 4.)  
Under this assumption we could also write the equivalent expression
\begin{equation}
\label{eq3.5}
\int_S E^\alpha \, dS_\alpha = 
\int_S \Etilde^\alpha \, \dStilde_\alpha =  - \frac{2}{3}q^2 
\int_{-\infty}^{\infty}
\left[ \frac{d\atilde^0}{d\tautilde}  + \atilde^2
 \utilde^0 \right] 
\, d\tautilde 
\q ,
\end{equation}
and this will be convenient for later comparison with the results of
\cite{hobbs2}.
It is customary to interpret the integrand 
\begin{equation} 
\label{eq3.7}
 - \frac{2}{3}q^2 
\left[ \frac{d\atilde^0}{d\tautilde}  + \atilde^2
 \utilde^0 \right] 
\q ,
\end{equation}
as the proper-time energy radiation rate.  (Interpreting 
$(2q^2/3) \atilde^2 \utilde^0 $ as the proper-time energy radiation
rate leads to an inconsistency: cf. \cite{parrott}, p. 140.) 
 
The interesting feature of \re{eq3} or \re{eq3.5} 
is that the acceleration which
appears in the expression for the radiated energy is the Minkowski
space acceleration $\atilde$ rather than the acceleration 
computed relative to the metric $g$.%
\footnote{The ``equivalence principle'' might lead one 
to expect the opposite.  However, this principle seems  
of dubious application to charged particles
\protect{\cite{parrott2}}.
}
Put another way, the energy radiation
is determined by the {\em apparent} 
acceleration relative to the static coordinate  frame rather than 
the physical acceleration experienced by the particle (i.e. 
the acceleration calculated using the semi-Riemannian connection
induced by $g$).  

This should not be surprising. 
For example, as previously noted, 
it seems universally believed 
that a stationary particle does not radiate energy 
even though a stationary particle is accelerated relative to $g$.
(Here and elsewhere we use terms like ``accelerated relative to $g$'' 
as shorthand for ``accelerated as measured by the
unique connection compatible with $g$.'')  

Now consider a charged particle which is stationary 
in the distant past, 
falls freely for a while, and then is
brought to rest and remains stationary thereafter.  
By ``falls freely'' we mean that it is in a state of zero proper
acceleration, so that its position $x^k (\tau)$ at proper time $\tau$ 
is governed by the geodesic equation,  
\begin{equation}
\label{geodesiceqn}
\frac{d^2{x^k}}{d\tau^2} = - \Gamma_{\alpha\beta}^k \frac{dx^\alpha}{d\tau}
\frac{dx^\beta}{d\tau} 
\end{equation}
where  $\Gamma_{ij}^k$ is the connection induced by $g$.

Suppose that the conformal factor $k$ 
in \re{eq1} depends only on $x^1$: $k = k(x^1)$ .
Then the nonzero connection coefficients are:
\begin{eqnarray*} 
\Gamma_{00}^1 &=&  \Gamma_{11}^1 = 
\frac{d\log k}{dx_1} =  
- \Gamma_{22}^1 = - \Gamma_{33}^1 \\ 
\Gamma_{01}^0 &=& \Gamma_{12}^2 = \Gamma_{13}^3 =  \frac{d\log k}{dx_1} 
\end{eqnarray*}
From this we see that a freely falling particle whose velocity
is initially in the $x_1$-direction will maintain constant
$x_2 , x_3$ coordinates forever,
and the $x^1$-component of \re{geodesiceqn} specializes to 
$$
\frac{d^2{x^1}}{d\tau^2} = - \frac{d\log k}{d x^1} 
\left[ \left(\frac{dx^0}{d\tau}\right)^2 + 
\left(\frac{dx^1}{d\tau}\right)^2  \right]
$$
This implies that when $d k /d x^1 \neq 0$, 
which we assume in this section,
the quantity $d^2 x^1/d\tau^2$ does not vanish 
during the period of free fall. 

The relationship between this quantity and the Minkowski space 
proper acceleration $\tilde{a}$ is messy, but one can see 
without detailed calculation that 
$\tilde{a}$ cannot vanish identically
unless $d^2 x^1/d\tau^2$ vanishes identically.
One way to see this is to note that at the start of the
free fall when the coordinate velocity $dx^1/dx_0$ vanishes, 
$$
\frac{d^2 x^1 }{ d\tau^2} = \frac{d~}{d\tau}
\left( \frac{dx^1}{dx_0}\frac{dx^0}{d\tau}\right)
= \frac{d^2x^1}{dx_0^2} \left( \frac{dx^0}{d\tau}\right)^2 
+ \frac{dx^1}{dx^0}\frac{d^2x^0}{d\tau^2} 
= \frac{1}{k^2} \frac{d^2x^1}{dx_0^2} 
$$
where the last equality is obtained by manipulating 
the identity $1 = k^2[(dx^0/d\tau)^2 - (dx^1/d\tau)^2]$
obtained from the definition \re{eq1} of the metric. 
Similarly, when the coordinate velocity vanishes,
the Minkowski space proper acceleration $\tilde{a}$
satisfies 
$$
\tilde{a}^2 = - (d^2x^1/dx_0^2)^2 = - k^4   (\frac{d^2 x^1 }{ d\tau^2})^2 \q 
$$
so that $\tilde{a}$ cannot vanish at this time.  

Given that $\tilde{a}$ does not vanish identically, 
the radiation given by \re{eq3} is nonzero because $\atilde^2 \leq 0$. 
To summarize, in a spacetime with metric \re{eq1} and $k = k(x^1)$
satisfying $dk/dx_1 \neq 0$, 
there {\em will} be energy radiation (as defined above) from
a charged particle initially at rest whose worldline thereafter 
is the same as that of a freely falling  
uncharged particle.

If energy is conserved, this implies that external forces would
have to be applied to drive a charged particle along such a worldline,
with the radiated energy supplied by these forces.
Put another way, if energy as defined above is to be conserved,
a charged particle acted on
by no external forces could not fall freely, contrary to the main 
result of \cite{hobbs2}.  Section 4 explores this situation in more detail. 

References \cite{dewitt/brehme} \cite{hobbs} \cite{hobbs2}
obtain their different results using, in effect, 
a different definition of
energy-momentum radiation which we question in Section 2. 
They obtain, in effect, expressions for
energy-momentum radiation in arbitrary spacetimes (not necessarily
static), but they express these in terms of equations of motion obtained
by setting up an equation of energy-momentum balance. 
When specialized to the case of the metric \re{eq1},
\cite{hobbs2} obtains the following equation of motion for
a particle of mass $m$ and charge $q$ in an external field $F$:
\begin{equation}
\label{eq4}
\label{hobbseq}
m \frac{du^i}{d\tau} 
= 
q{F^i}_\alpha u^\alpha + 
\frac{2}{3}q^2 \left[ \frac{da^i}{d\tau} + 
a^2 u^i \right] 
- \frac{1}{3} q^2 \left[ - {R_\beta}^i u^\beta 
+  u^i R_{\alpha \beta} u^\alpha u^\beta \right] .
\end{equation}
Here $\tau$ is proper time, $\tau \mapsto z(\tau )$ the particle's
worldline, $u := dz/d\tau$ its four-velocity, 
$a := du/d\tau$ its proper acceleration, and $R$ the Ricci tensor.
The difference in certain signs between \re{hobbseq} and equation 
(5.28) of \cite{hobbs} is because \cite{hobbs} uses a metric of signature
$(-1,1,1,1)$ opposite to ours. 

The left side $m(du^i / d\tau)$ represents the rate of change
of mechanical energy-momentum of the particle.  The first term on
the right, $q{F^i}_\alpha u^\alpha$, is the rate at which 
the external field furnishes energy-momentum.  The remaining terms
on the right represent the negative of the energy-momentum radiation
rate.

This differs from our expression \re{eq3} for energy radiation 
in a several essential ways. 
First of all, the bracketed last term in \re{hobbseq} 
involving the Ricci tensor has no counterpart in our \re{eq3}.   
However, \cite{hobbs2} notes that the term involving the Ricci tensor
vanishes for certain special conformal factors $k$.  One such is 
\begin{equation}
\label{eq5.5}  
k(x^0, x^1, x^2, x^3 ) =  \frac{1}{x^1} 
\q ,
\end{equation}
and we shall use this as a test case to compare the two approaches.

It is not entirely clear what is the proper way to compare \re{eq4}
with our results \re{eq3} or \re{eq3.7}, but it does not seem likely 
that the conclusions of \cite{hobbs2} can be easily reconciled 
with ours. 
Since for certain conformal factors 
the term involving the Ricci tensor vanishes,
the important radiation term in \re{eq4} would seem to be 
\begin{equation}
\label{eq5.2}
\frac{2}{3}q^2 \left[ \frac{da^i}{d\tau} + a^2 u^i \right]
\q .
\end{equation}
In Minkowski space, this is just the radiation term in the Lorentz-Dirac
equation. 

If we interpret \re{eq5.2} as the proper-time energy-momentum radiation
rate, then it would seem plausible to interpret the inner product
of \re{eq5.2} with the unit vector $k^{-1}\pl_0$ as the proper-time
energy radiation rate, the energy being calculated relative to the
coordinate frame.  
Under this interpretation, the Hobbs method gives
an energy radiation rate of 
\begin{equation}
\label{eq5.3} 
\frac{2}{3}q^2 k \left[ \frac{da^0}{d\tau} + a^2 u^0 \right] = 
\frac{2}{3}q^2 k \left[ \frac{da^0}{d\tau} + a^2 k^{-1} \utilde^0 \right]  
\q ,
\end{equation}
for the above case \re{eq5.5} in which the Ricci term vanishes.
The factors of $k$ are of no significance here.  We are concerned with
the proper-time radiation rate at a particular point on the worldline,
and $k$ can be normalized to unity at this point.  The really significant
difference between \re{eq5.3} and \re{eq3.7} is the replacement of the
apparent (or Minkowski) acceleration $\atilde$ in \re{eq3.7} by the
proper acceleration $a$ in \re{eq5.3}.  Even if one questions the above 
interpretation, it is clear that this fundamental difference between
our expressions and those of \cite{hobbs2} remains.
In the face of such inconsistency, it is
appropriate to examine both methods for possible sources of error.

\section{Objections to the DeWitt/Brehme/Hobbs method.} 

It seems to us that the most questionable feature of the 
method of \cite{dewitt/brehme} and \cite{hobbs}
is that its derivation of the equation of motion of a charged 
particle in arbitrary spacetimes
employs a physically unjustified identification of
all tangent spaces in the neighborhood of a point on the particle's
worldline in order to calculate (in effect) the radiated energy-momentum.  

In outline, their method is as follows.  Surround the
particle with a small two-dimensional sphere $S_\tau$ associated with a given proper
time $\tau$ on the worldline.  
There are several reasonable ways to construct such spheres, and there
is no reason to think that the final equation of motion will be
independent of the method of construction (see below).
However, let us pass over this point for the moment.

As this two-dimensional sphere evolves
through time, it generates a three-dimensional ``tube'' $\Sigma$ surrounding
the worldline.  In Minkowski space, the integral 
\begin{equation}
\label{tubeint}
P^i := \int_{\Sigma} T^{i\alpha} \, d\Sigma_\alpha 
\end{equation}
of the electromagnetic
energy-momentum tensor $T$ over this tube yields a vector quantity $P^i$
which is physically interpreted as the energy-momentum radiated by the 
particle.  The equation of motion is then obtained as as an equation of
energy-momentum balance.  This is the method by which Dirac obtained
the Lorentz-Dirac equation \cite{dirac} for a charged particle in
Minkowski space.  

A fundamental difficulty in attempting to use the same method 
in an arbitrary spacetime, is that the integral \re{tubeint}
is not well-defined, because in effect it attempts the illegitimate
mathematical operation of summing vectors in different 
tangent spaces. 
In Minkowski space this difficulty does not arise because the linear structure
of the space gives natural identifications of all tangent spaces.  
To generalize \re{tubeint} to arbitrary spacetimes, one needs some replacement
for this natural identification.
%\footnote{In response to a concern raised in a private communication from
%one of the cited authors, we want to make clear that both \cite{dewitt/brehme}
%and \cite{hobbs} recognize this.}

Both \cite{dewitt/brehme} and \cite{hobbs} do recognize this difficulty, 
but the identifications they use 
are introduced without physical motivation.
They seem arbitrary, and there seems to be no general principle guaranteeing
that other, equally reasonable identifications would yield the same
equation of motion.  To illustrate, consider the following
three methods.   
\begin{description}
\item[Method 1:]   
Choose a ``base'' point $z(\tau_0 ) $ 
on the worldline $\tau \mapsto z(\tau )$.  
Any point $x$ in a sufficiently small neighborhood of $z(\tau_0 )$ can 
be joined to $z(\tau_0 )$ by a unique geodesic lying in this neighborhood.  
Parallel translation
along this geodesic yields an identification of the tangent space at
$x$ with the tangent space at $z(\tau_0 )$.  In this way, all tangent
spaces in a sufficiently small neighborhood of $z(\tau_0 )$ are
identified.  
\\[1ex]
This identification would be expected to depend
on the base point when the curvature does not vanish, 
since the difference
between the identifications defined by two base points is parallel translation
around a geodesic quadrilateral.   
Nevertheless, one could in principle 
use the method of \cite{dewitt/brehme} to obtain an equation of motion 
using this identification.   

The dependence of the identification on the base point does expose 
the method to certain criticisms.
Similar criticisms given below apply to the identifications
actually used in \cite{dewitt/brehme} and \cite{hobbs}. 
\item[Method 2:]  Again choose a base point $z(\tau_0 )$ on the
worldline.  From an arbitrary point $x$ near $z(\tau_0 )$, find a 
geodesic $\gamma $ connecting $x$ to a point $z(\tau_1 )$ on the
worldline with the property that at $z(\tau_1 )$, $\gamma$ is orthogonal
to the worldline.  (In Minkowski space, $\gamma$ would be a straight line
orthogonal to the worldline, as pictured in Figure \ref{worldline}.)  
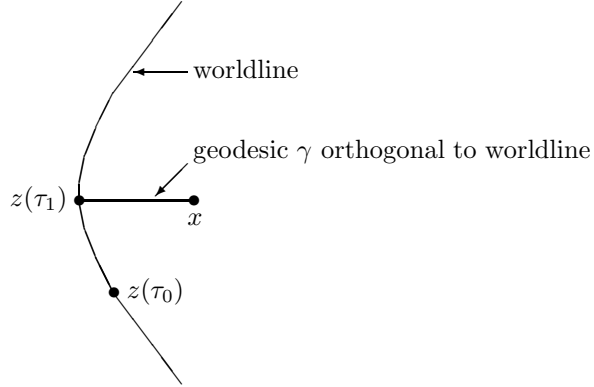
\begin{figure}
\vspace{20mm}
\begin{picture}(4,4)(0,-2)
\setlength{\unitlength}{.6in}
\put (1.0,-.07) {\line(0,1){.14}}
%draw orthogonal geodesic and various labels
\put (1.0,-.07){\line(1,0){1}}
\put (.95,-.13){$\bullet$}
\put (1.95,-.13){$\bullet$}
\put (2, 1){worldline}
\put (1.95, 1.05){\vector(-1,0){.47}}
\put (2, .3){geodesic $\gamma$ orthogonal to worldline}
\put (1.95, .25){\vector(-1,-1){.28}}
\put (1.95, -.28){$x$}
\put (.4,-.12){$z(\tau_1)$}
%end orthogonal geodesic
\put (1.0,.07) {\line(1,5){.05}}
\put (1.05,.32) {\line(2,5){.1}}
\put (1.15,.57){\line(1,2){.15}}
\put (1.3,.87){\line(3,4){.6}}
%\put (1.9,1.67){\line(4,5){.2}}
\put (1.0,-.07) {\line(1,-5){.05}}
\put (1.05,-.32) {\line(2,-5){.1}}
\put (1.15,-.57){\line(1,-2){.15}}
\put (1.3,-.87){\line(3,-4){.6}}
\put (1.25,-.93){$\bullet$}
\put (1.42,-.92) {$z(\tau_0 )$}
\put (1.15,-.57){\line(1,-2){.15}}
%\put (1.9,-1.67){\line(4,-5){.2}}
\end{picture}
%\vspace{40mm}
\vspace{30mm}
\caption{The tangent space at a point $x$ near the ``base'' point
$z(\tau_0)$ is identified with the tangent space at $z(\tau_0)$
via parallel transport on the unique geodesic starting at $x$
and orthogonal to the worldline followed by Fermi-Walker transport
from the end of that geodesic at $z(\tau_1 )$ to $z(\tau_0 )$.} 
\label{worldline}
\end{figure}
Identify the tangent space at $x$ with the 
tangent space at $z(\tau_0 )$ by using parallel translation along
$\gamma$ from $x$ to $z(\tau_1 )$ and then Fermi-Walker transport 
along the worldline  from $z(\tau_1 ) $ to $z(\tau_0 )$.  
\\[1ex]
This identification does not depend on the base point $z(\tau_0 )$,
but does have the more subtle defect of depending on the
worldline.  When one computes the integral \re{tubeint}
using this identification, it is not clear what is the physical
meaning of the ``vector'' $P^i$ obtained.  For instance, if we
imagine performing the integration for two different worldlines 
through the same base point $z(\tau_0 )$, we obtain two different 
$P^i$'s, and there would seem to be no sensible way to compare 
them.  

Such a comparison is not necessary for the derivations
of \cite{dewitt/brehme} and \cite{hobbs}, but if \re{tubeint} is
computing some real physical quantity, such a comparison should
be possible.  For instance, the rate of change of mechanical
energy-momentum $m(du/d\tau )$ on the left of \re{eq4} can be
sensibly compared for two worldlines, since these are tangent 
vectors at the same point.  
If the left sides of \re{eq4} can be
sensibly compared, then one would think that it should be possible to
compare the $P^i$'s on the right. 

The difficulty can be seen more clearly by imagining 
a collision process in which the worldlines of two particles  
intersect at one point.  For such a situation, it would seem reasonable
to account for the interchange of mechanical energy-momentum
by simply summing energy-momentum vectors of the incoming and outgoing
worldlines, but the method of \cite{dewitt/brehme} and \cite{hobbs}
would not be expected to correctly account for radiation in this context.

This is a situation which they do not consider, and the inapplicability
of their method is not in itself a compelling objection.  It could be that 
point collision processes are inherently unrealistic and that no 
method could properly account for them.   However, it does seem a
reason to closely scrutinize the method.
\item[Method 3:]  This is a variant of Method 2.  Instead of choosing 
$\gamma$  orthogonal to the worldline, choose $\gamma$ to be a lightlike
geodesic from $x$ to the worldline.   
\end{description}
An advantage of Methods 2 and 3 is that one needs to consider only
spacelike or only lightlike geodesics, which sometimes leads to 
calculational simplifications. 
Hobbs (\cite{hobbs}, p. 145) uses Method 2, though his calculational
method effectively bypasses the Fermi-Walker part of the translation.

Methods 2 and 3 also give more or less ``natural'' ways to construct
the spheres $S_{\tau}$.  For example, with Method 2, $S_{\tau_1}$
could be taken as the set of all $x$ whose connecting geodesic to
$z(\tau_1 )$ has a fixed length $r$.  For Method 3, one could 
replace the geodesic length by the geodesic parameter, under the
normalization condition that the inner product of the geodesic
tangent $\gamma\pr$ with the worldline tangent $z\pr (\tau_1 )$ be unity.

In Minkowski space, these last two constructions yield tubes called 
Dirac and Bhabha tubes, respectively, and both calculations can
be done exactly for the limit of a tube of vanishing
radius.  The integrals \re{tubeint} over the two
tubes between finite times $\tau_1$ and $\tau_2$ do {\em not} coincide 
(\cite{parrott}, p. 160), though they are close enough that a plausible
argument can be made for the Lorentz-Dirac equation.  The situation in
arbitrary spacetimes seems much more obscure.  We know of no good reason
to think that any two of the three methods (or others equally ``natural'')
will yield equivalent results.  
 
Another objection to the method of \cite{dewitt/brehme} and \cite{hobbs} 
is that certain 
(probably divergent) integrals are discarded with little discussion
or physical justification.  These discarded integrals are the
integrals over the spacelike 3-volumes $\Sigma_1$ and $\Sigma_2$
and the integral over the 4-volume $V$ in equation (5.2) of 
\cite{hobbs} and the similar integrals in equation (5.1) of 
\cite{dewitt/brehme}.  The integrals over the spacelike 3-volumes
(which may be roughly visualized as constant-time hypersurfaces)
would yield mass renormalization terms in Minkowski space, 
assuming that the particle is unaccelerated in a neighborhood 
of the hypersurfaces. 
When the particle is accelerated at the hypersurfaces (as it
would be in general in the formulations of \cite{dewitt/brehme}
and \cite{hobbs}), the corresponding integrals have never been
computed even in Minkowski space, and we know of no reason to
believe that they will evaluate to mass renormalizations.

\section{Critical discussion of our method.}

Now let us look for possible sources of error in our expression 
\re{eq3} for the energy radiation.  It is customary in the 
relativity literature to identify as ``energy'' the conserved quantity
corresponding to $\partial_0$ in a static spacetime.  Nevertheless,
we have pointed out in \cite{parrott2} that such an identification
is occasionally physically incorrect.  For example, for the
metric
\begin{equation}
\label{eq9}
g^{ij} dx^i dx^j  = (x^1)^2 (dx^0)^2 - \sum_{I=1}^3 (dx^I)^2
\q ,
\end{equation}
the Riemann tensor vanishes, which implies that this spacetime may
be metrically identified with a subset of Minkowski space, whose
metric is
\begin{equation}
\label {eq9.5}
ds^2 = dt^2 - dx^2 - dy^2 - dz^2
\q .
\end{equation}
In Minkowski space, the energy is universally accepted as the conserved
quantity associated with the Killing vector $\partial_t$, and this
is not the same as the conserved quantity associated with
$\partial_0 $.  

If there are timelike Killing vectors other than $\partial_0$ 
for our metric \re{eq1}, it is conceivable that in analogy to
the situation just presented, one of these might give the physically
relevant energy as conserved quantity rather than $\partial_0$.
However, it is rather rare that such hidden symmetries exist,
and we show in the Appendix 
that for the metric \re{eq1}
with $k$ given by \re{eq5.5}, the only Killing
vector fields which are invariant under translations and rotations
in the $x^2$-$x^3$ plane are constant multiples of $\pl_0$.  Thus
apart from a trivial multiplicative constant, $\pl_0$ seems the only
natural choice for the Killing field associated with ``energy''. 
\section{A physical argument suggesting that the equation
of motion \protect{\re{eq4}} may be incorrect.}
This section argues that if 
we do identify $\partial_0$ as the ``energy'' Killing vector,
then it appears unlikely that Hobbs' equation of motion \re{eq4} can be correct
for a conformally flat spacetime \re{eq1} with $k$ given by
\re{eq5.5}.  

It is well known that on any geodesic with tangent vector $u$, 
the inner product $\langle u, \partial_0  \rangle $  
is constant (\cite{mtw}, p. 651).
In the present situation, this says that the four-velocity $u$ of a freely
falling particle satisfies 
\begin{equation}
\label{eq7}
\langle u , \partial_0 \rangle = \frac{k}{\sqrt{1-v^2}} = \mbox{constant} 
\q ,
\end{equation}
where $v^2 := \sum_{J=1}^3 (dx^J / dx^0)^2 $ is the square of the coordinate
velocity.  To see this, note that $ u = (dx^0 /d\tau , d{\bf x } /d\tau ) = 
(dx^0 / d\tau )(1, d {\bf x }/dx^0 )$,
where $\tau$ is proper time.  It follows that  
$1 = u^2 = k^2 (dx^0 / d\tau )^2 (1 - v^2)$,
so that $\la u, \pl_0 \ra = k^2 dx^0 / d\tau = 
k/ \sqrt{1-v^2}$.     

The logarithm of equation \re{eq7} may be regarded as a sort of law of
conservation of kinetic plus potential energy: $\log k $ may be regarded 
as the potential
energy.  For example, 
if $k$ is decreasing as the particle moves along a geodesic, 
then $v$ is increasing.  In particular, a freely falling 
particle which is released at a ``height''
$ k = k_0$ with a particular
velocity will arrive at a lower ``height'' $k = k_1 < k_0$ 
with a greater velocity.  

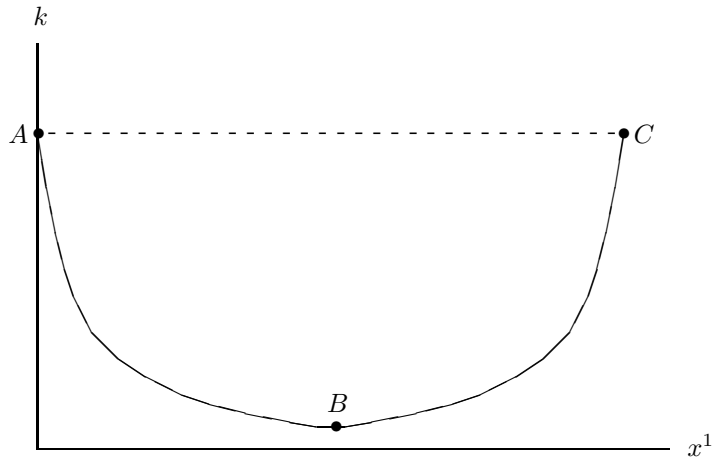
\begin{figure}
%\vspace{20mm} 
\vspace{25mm} 
%\begin{picture}(90,70)(-5,-10)
\begin{picture}(90,70)(-10,-10)
\setlength{\unitlength}{1.2mm}
%draw axes and label
\put (0,-5) {\line(1,0){70}}
\put (72,-5.7) {$x^1$} 
\put (0,-5) {\line(0,1){45}}
\put (-.4,42) {$k$}
%draw curve
\put (0,30) {\line(1,-6){1}}
\put (1,24) {\line(1,-5){1}}
\put (2,19) {\line(1,-4){1}}
\put (3,15) {\line(1,-3){1}}
\put (4,12) {\line(1,-2){2}}
\put (6,8) {\line(1,-1){3}}
\put (9,5) {\line(3,-2){3}}
\put (12,3) {\line(2,-1){4}}
\put (16,1) {\line(3,-1){3}}
\put (19,0) {\line(4,-1){4}}
\put (23,-1) {\line(5,-1){5}}
\put (28,-2) {\line(6,-1){3}}
\put (31,-2.5) {\line(1,0){3}}
\put (34,-2.5) {\line(6,1){3}}
\put (37,-2) {\line(5,1){5}}
\put (42,-1) {\line(4,1){4}}
\put (46,0) {\line(3,1){3}}
\put (49,1) {\line(2,1){4}}
\put (53,3) {\line(3,2){3}}
\put (56,5) {\line(1,1){3}}
\put (59,8) {\line(1,2){2}}
\put (61,12) {\line(1,3){1}}
\put (62,15) {\line(1,4){1}}
\put (63,19) {\line(1,5){1}}
\put (64,24) {\line(1,6){1}}
%label curve
\put (32.4,-3.2) {$\bullet$}
\put (32.1, -.9)  {$B$}
\put (-.6,29.3) {$\bullet$}
\put (-3.2,29) {$A$}
\put (64.25,29.3) {$\bullet$}
\put (66.1,29) {$C$}
%Draw horizontal dashed line
\multiput(1.2,29.45)(2,0){32}{- }
\end{picture}
\vspace{5mm}
\caption{A freely falling particle oscillates
forever on the $x^1$-axis between the projections of $A$ and $C$ to
the $x^1$-axis.}
\label{fig2}
\end{figure}

If we had the potential energy given by a function 
$k = k(x^1)$ of the form sketched in 
Figure \ref{fig2},
then a freely falling particle released 
at rest at point $A$ would fall to arrive 
with nonzero velocity at the point $B$ of minimum potential energy.%
\footnote{
Since the motion is on the $x^1$-axis, the particle 
actually moves not from $A$ to $B$, but rather from the point on the
horizontal axis whose $x^1$ coordinate is that of $A$ 
to one whose $x^1$ coordinate
is $B$. However, speaking of ``falling'' from $A$ to $B$ allows us 
to describe the motion in physically suggestive language 
and should cause no confusion.
}  
By symmetry, it would then climb to arrive at $C$ with zero velocity,
after which it would fall back toward $B$.  The oscillations
$A$-$B$-$C$-$B$-$A$ would continue forever.

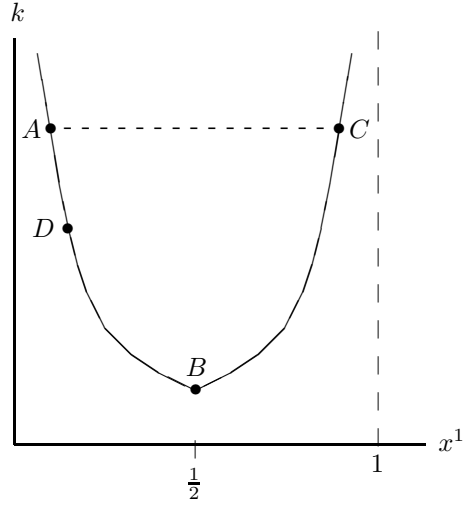
\begin{figure}
%Figure 3
\setlength{\unitlength}{1.2mm}
\begin{picture}(60,55)(-15,-15)
\setlength{\unitlength}{1.2mm}
%draw axes and label
\put (-4,-5) {\line(1,0){45.5}}
\put (43,-5.7) {$x^1$} 
\put (15, -10) {$\frac{1}{2}$}
\put (15.7, -6) {$\scriptstyle |$}
\multiput (36,-4.7)(0,4){12}{$\scriptstyle|$} 
\put (35.5,-8){$1$}
\put (-4,-5) {\line(0,1){45}}
\put (-4.4,42) {$k$}
%draw curves and label
\put (-1.4,38.4) {\line(1,-6){1.4}}
\put (0,30) {\line(1,-6){1}}
\put (1,24) {\line(1,-5){1}}
\put (2,19) {\line(1,-4){1}}
\put (1.2,18.3) {$\bullet$}
\put (-2.0,18) {$D$}
\put (3,15) {\line(1,-3){1}}
\put (4,12) {\line(1,-2){2}}
\put (6,8) {\line(1,-1){3}}
\put (9,5) {\line(3,-2){3}}
\put (12,3) {\line(2,-1){4}}
\put (16,1) {\line(2,1){4}}
\put (20,3) {\line(3,2){3}}
\put (23,5) {\line(1,1){3}}
\put (26,8) {\line(1,2){2}}
\put (28,12) {\line(1,3){1}}
\put (29,15) {\line(1,4){1}}
\put (30,19) {\line(1,5){1}}
\put (31,24) {\line(1,6){1}}
\put (32,30) {\line(1,6){1.4}}
%label curve
\put (15.4,.4) {$\bullet$}
\put (14.98, 2.6)  {$B$}
%\put (-.6,29.3) {$\bullet$}
\put (-.7,29.3) {$\bullet$}
\put (-3.2,29) {$A$}
\put (31.25,29.3) {$\bullet$}
\put (33.1,29) {$C$}
%Draw horizontal dashed line
\multiput(1.5,29.45)(2,0){16}{- }
\end{picture}
\vspace{-5mm}
\caption{The ``potential'' function $k$ is defined by  $k(x^1) := 1/x^1$
\protect{\newline} 
for $0 \leq x^1 \leq 1/2$ and 
$k(x^1) := 1/(1-x^1)$ for $1/2 \leq x^1 \leq 1$.  The sharp corner at $B$ 
can be rounded off, if desired, leaving a small transition region 
near $B$ where the bracketed term involving the Ricci tensor in
\protect{\re{eq4}} does not vanish.
}
\label{fig3}
\end{figure}

If a charged particle is radiating energy during this motion, we would
have an infinite energy source.  We are saved from a direct paradox
by the fact that the graph of $k(x^1) = 1/x^1$ does not have 
a local minimum,
but we can still obtain a result which is physically hard to believe by
putting two similar $k$'s together as sketched
in Figure \ref{fig3}.
The sharp corner at the minimum at $x^1 = 1/2$ can be rounded off, if 
desired, leaving a small transition region near $B$ where the
bracketed term involving the Ricci tensor in \re{eq4} does not
vanish.  

If a particle is released at rest at point $ A$, the
DeWitt/Brehme/Hobbs equation \re{eq4} implies 
that it will fall freely until
it reaches the transition region near point $B$.  
Our energy calulation indicates that 
it is radiating energy during this period.  The energy radiated at
a point $D$ intermediate between $A$ and $B$ can presumably be collected by 
antennas located near $D$.  If the particle oscillated indefinitely
between $A$ and $C$, we would have an infinite energy source, so
conservation of energy requires that 
the particle lose velocity (relative to the freely falling
motion of an otherwise identical uncharged particle) in the transition
region near $B$.

It is conceivable that the energy collected by antennas near points 
like $D$ where the particle is freely falling 
might be ``borrowed'' until the particle reaches 
the transition region near $B$, the decrease in velocity over the transition
region being precisely that required to pay for the radiated energy.   
However, since there is
nothing in the structure of the DeWitt/Brehme/Hobbs equation to guarantee
this, it seems unlikely.  Even if the equation did guarantee it, 
such ``borrowing'' of energy,
though not a mathematical contradiction, seems physically very strange.
We think it suggests that the DeWitt/Brehme/Hobbs equation should not
be uncritically accepted. 

%Thus for this choice of $k$, the
%only reasonable choice for the Killing field yielding ``energy''
%is (apart from a trivial constant) the obvious choice $ \partial_0$.
%If one insists in believing in the equation of motion \re{eq4} for
%this $k$, then the only way out of the impasse would seem to be
%to deny that the conserved quantity corresponding to 
%$\partial_0$ is ``energy'', in which case there would be no 
%conserved ``energy'' at all, or at least none constructed from
%a Killing vector.  
%However, if we can't identify $\partial_0$ with energy
%in this situation, then we should presumably look with suspicion
%on the many other situations in general relativity where similar
%identifications are routinely made.
\section{Appendix}
This appendix classifies the Killing vector fields $K = K^i \pl_i$ 
for a metric of
the form 
\re{eq1} with $k$ given by \re{eq5.5}:  
\begin{equation}
\label{eq8}
g_{ij} dx^i dx^j = (1/x^1)^2 [ (dx^0)^2 - \sum_{J=1}^3 (dx^J)^2 ]
\q .
\end{equation}
The result is that any such Killing field is of the form
\begin{equation}
\label{eq8.5}
K = K^0 ( x^2 , x^3 ) \pl_0  + 
K^2 (x^0 , x^3 ) \pl_2 + K^3 
(x^0 , x^2 , ) \pl_3
\q ,
\end{equation}

In other words, any Killing field for \re{eq8} is actually an $x^1$-independent
Killing field for the three-dimensional Minkowski metric 
$(dx^0)^2 - (dx^2)^2 - (dx^3)^2$.  
\footnote{The latter Killing fields are classified 
in \cite{weinberg}, Chapter 13, 
under the additional hypothesis that the 
$K^i$ be analytic.}
Moreover, it will be apparent from
the Killing equation \re{eq10} below that the space part  
$K^2 (x^0 , x^3 ) \pl_2 + K^3  (x^0 , x^2 , ) \pl_3 $ of $K$ is 
an $x^0$-dependent Killing vector for the Euclidean $x^2$-$x^3$ plane.
The Killing fields for a Euclidean plane are well-known to be linear
combinations (with constant coefficients) 
of constant vector fields and the infinitesimal generator 
of the rotation group; 
it follows that  the only Killing fields for \re{eq8} which are invariant
under translations and rotations in the $x^2$-$x^3$ plane are constant
scalar multiples of $\pl_0$.  
%Since such invariance is a natural requirement
%to impose on a Killing field to represent ``energy'', this indicates 
%that the only natural choice for an ``energy'' Killing field 
%is a nonzero scalar multiple of the obvious one $\pl_0$.  

The condition that $K$ be a Killing field may be written as
(\cite{sachs/wu}, p. 81, Problem 3.6.3(b)):
\begin{equation}
\label{eq10}
g_{i\alpha} \pl_j K^\alpha + g_{j\beta} \pl_i K^\beta + K(g_{ij})=0
\q .
\end{equation}
Here we use the differential-geometric convention of identifying 
tangent vectors with directional derivatives, so that 
$K(g_{ij}) := K^\alpha \pl_\alpha g_{ij}$.

For a diagonal metric like \re{eq8}, this simplifies to
\begin{equation}
\label{eq11}
g_{ii}\pl_j K^i + g_{jj} \pl_i K^j + K(g_{ij}) = 0 \q \q
\mbox{(NO SUMS)},
\end{equation}
where the repeated indices $i$ and $j$ are {\em not} summed.
%We first consider this equation for the case in which $i$ and $j$
%are both drawn from the set $\{ 0,1\}$, and we'll show that it
%can hold only if $K^1 = 0$.  Assuming this, the fact that $g_{ij}$
%depends only on $x^1$ implies that 
We shall show that \re{eq11} 
can hold only if $K^1 = 0$.  Assuming this, the fact that $g_{ij}$
depends only on $x^1$ implies that 
\begin{equation}
\label{eq11.5}
K(g_{ij}) = K^1 \pl_1  g_{ij} = 0
\q .
\end{equation}
From this and \re{eq11}, it follows that 
$K^i$ is independent of both $x^1$ and $x^i$, establishing \re{eq8.5}. 

To show that $K^1 = 0$, we consider \re{eq1} for 
the special case in which $i$ and $j$
are both drawn from the set $\{ 0,1\}$.  
While considering this special case, we simplify the notation by suppressing 
the $x^2 , x^3$ dependence of $K^i$.   Let $ k(x^1)  := 1/x^1$.  (Our
argument actually works for a large class of $k$'s.)

From \re{eq11} for $i=j=1$, 
we obtain
$$
0 = -2k^2 \pl_1 K^1 - K(k^2 ) = -2 k^2 \pl_1 K^1 - K^1 \pl_1 (k^2) 
\q .
$$
If $K^1 \neq 0$ at some point, we may divide by $K^1$, obtaining
near that point
\begin{equation}
\label{eq12}
\pl_1 \log K^1 = - \pl_1 \log k
\q . 
\end{equation}
We'll show that this leads to a contradiction, so that the only
alternative is the promised relation $K^1 = 0$, from which the
rest of the conclusions follow routinely as outlined above.

From  \re{eq12}, it follows that  
\begin{equation}
\label{eq13}
K^1 (x^0 , x^1 ) = \phi (x^0 )/k(x^1) 
\q 
\end{equation}
for some function $\phi$ of $x^0 $ (and the suppressed $x^2$ and $x^3$) alone.
Next write \re{eq11} for $i=j=0$ and again for $i=j=1$ to conclude that 
$$
2k^2 \pl_0 K^0 = -K(k^2 ) = 2k^2 \pl_1 K^1
\q ,
$$
whence 
\begin{equation} 
\label{eq14}
\pl_0 K^0 = \pl_1 K^1
\q .
\end{equation}
Finally, write \re{eq11} for $ i=0 , j=1$ to see that
\begin{equation}
\label{eq15}
\pl_1 K^0 = \pl_0 K^1
\q .
\end{equation}
Applying $\pl_0$ to \re{eq15} and using \re{eq14} shows that
$ K^1$ satisfies the wave equation 
$$\pl_0^2 K^1 = \pl_1^2 K^1 \q .$$ 
But a function of form \re{eq13} can satisfy 
the wave equation only if 
$1/k$ satisfies an equation of the form
$(1/k)^{\prime \prime} + \lambda (1/k) = 0 $ for some constant $\lambda$; i.e. 
$1/k(x^1) $ must be a linear combination of complex exponentials 
$e^{\sqrt{\lambda} x^1}$.
Since our $k(x^1) := 1/x^1$ is not of this form, we conclude that 
\re{eq11} can hold only for $K^1 = 0$.

Finally, from \re{eq10} and \re{eq11.5}, it follows that 
$K^2(x^0, x^3) \pl_2 + K^3 (x^0, x^2) \pl_3$ is an $x^0$-dependent
Killing field on the $x^2$-$x^3$ plane.  As noted above, the only such
field invariant under translations and rotations is the zero field. 
\section{Appendix 2.} 
This appendix describes significant developments which have
occurred since this paper was written in late 1995 
and which affect some of its arguments.
The first is a theorem of Sonego \cite{sonego} showing the 
conformal invariance 
of the DeWitt/Brehme/Hobbs (abbreviated DBH below) 
radiation reaction term in equation \re{hobbseq}, 
\begin{equation}
\label{hobbsrad}
\frac{2}{3}q^2 \left[ \frac{da^i}{d\tau} + 
a^2 u^i \right] 
- \frac{1}{3} q^2 \left[ - {R_\beta}^i u^\beta 
+  u^i R_{\alpha \beta} u^\alpha u^\beta \right] 
\q.
\end{equation}
The second is a long paper of Quinn and Wald \cite{quinn/wald}
relating the ``Killing vector'' definition of energy radiation
\re{eq2.5} to the DBH equation.
These imply that the plausibility argument of Section 4
is basically a disguised version of a known
physical objection to the Lorentz-Dirac equation 
discussed in detail in \cite{parrott2} and
\cite{parrott3} and outlined below. 

\subsection{Implications of conformal invariance of the DBH radiation reaction}

Sonego proved that the radiation-reaction term $\phi^i$ 
in the DBH equation \re{hobbseq},
\begin{equation}
\label{eqA1}
\phi^i := 
\frac{2}{3}q^2 \left[ \frac{da^i}{d\tau} + a^2 u^i \right] 
- \frac{1}{3} q^2 \left[ - {R_\beta}^i u^\beta 
+  u^i R_{\alpha \beta} u^\alpha u^\beta \right] 
,
\end{equation}
is conformally invariant in the sense that
if metrics $g$ and $\tilde{g}$ are related by $ g = k^2 \tilde{g}$
for some $C^\infty$ function $k$, then 
\begin{equation}
\label{eqA2}
 \phi^i = k  \tilde{\phi^i} 
\q,
\end{equation}
where $ \tilde{\phi^i}$ is defined by the right side of \re{eqA1}
with all quantities computed relative to $\tilde{g}$ 
instead of $g$. 
This does not require that $\tilde{g}$ be the Minkowski metric
of the present work, but this is the only case we consider here.

For the $g$-geodesics considered in Section 4, by definition $a = 0$. 
Also, the term in \re{eqA1}
involving the Ricci tensor vanishes. 
Hence Sonego's result implies that 
\begin{equation} 
\label{eqA3} 
 \frac{d\tilde{a}^i}{d\tilde{\tau}} + \tilde{a}^2 \tilde{u}^i = 0 
\q.
\end{equation}
This implies that the $g$-geodesics are {\em uniformly accelerated}
relative to the Minkowski metric $\tilde{g}$.  

Indeed, this is one way to formulate the definition of ``uniform acceleration'';
after observing that $\atilde^2 = - \langle d\atilde/d\taut , \utilde \rangle$,
we see that
it states that $\tilde{a}$ is invariant under Fermi-Walker transport
along the worldline.  
Alternatively, the identification
of \re{eqA3} with uniform acceleration for the present situation
of one-dimensional motion can be established more directly 
by writing $\tilde{a} = \tilde{A}\tilde{w}$ with $\tilde{w}$ a unit
vector orthogonal to $\tilde{u}$ and $\tilde{A}$ the scalar proper
acceleration.  Noting that $d\tilde{w}/d\tilde{\tau}$ is orthogonal
to $\tilde{w}$ (because $\tilde{w}$ is a unit vector) and
extracting the $\tilde{w}$ component of 
$$
\frac{d\tilde{a}}{d\tilde{\tau}}    
= \frac{d\tilde{A}}{d\tilde{\tau}} \tilde{w} + 
\tilde{A} \frac{d\tilde{w}}{d\tilde{\tau}} 
$$
shows that \re{eqA3} implies (actually, is equivalent to) 
$d\tilde{A}/d\tilde{\tau} = 0$, 
i.e., constant scalar proper acceleration. 

This establishes a connection between geodesic motion 
in the $g$-spacetime and uniformly accelerated motion
in Minkowski space.%
\footnote{The argument just given applies more generally 
to correspond to any uniformly accelerated worldline
in the $g$-spacetime (of which geodesic motion is an instance), 
a uniformly accelerated worldline in Minkowski space.
However, we do not need this.} 
We will say more about this connection later. 
But first we review some strange consequences  of the assumption
that the Lorentz-Dirac radiation reaction term  correctly
describes radiation reaction for a uniformly accelerated particle
in Minkowski space.  A more extensive discussion with full mathematical
details can be found in \cite{parrott2}. 

Suppose a rocket ship contains a charged particle as payload.
The ship also contains some additional mass which 
serves as fuel, by converting mass to energy.

Suppose the ship is unaccelerated in Minkowski space
for all time up to some initial time $\tilde{\tau}_i$.  
At that time, it starts its engines and makes a smooth transition
to a uniformly accelerated state at a slightly later time
$\tilde{\tau_i} + \epsilon$. 
(All motion is in one spatial dimension.) 
It continues its uniform acceleration for a long time,
finally smoothly removing the uniform acceleration during a 
transition interval $[\tilde{\tau}_f - \epsilon , \tilde{\tau}_f]$.
after which it is unaccelerated again.  
In summary, it has a smooth worldline which  
is locally uniformly accelerated
(with zero acceleration before the initial time $\taut_i$
and after the final time $\taut_f$)
except during the transition
intervals $[\taut_i, \taut_i + \epsilon]$ and 
$[\taut_f - \epsilon, \taut_f]$. 
It is assumed that the worldline is $C^\infty$; 
the function of the transition intervals is to permit a smooth 
transition from inertial motion (zero acceleration) to constant
and nonzero uniform acceleration. 

We will refer to this situation 
as ``uniform acceleration for a finite time''.  
We emphasize that 
essential features of ``uniform acceleration for a finite time''
are absence of acceleration in distant past and future% 
\footnote{Asymptotically vanishing acceleration
in distant past and future might be good enough for some applications,
but the additional generality thus obtained usually is insufficient
compensation for the difficulty of carrying out the arguments
with mathematical rigor.  The problem is that
one can easily obtain demonstrably incorrect results from plausible
manipulations if one ignores conditions at the initial and 
final times $\taut_i$ and $\taut_f$.   
Both zero acceleration and smoothness of the worldline at these times
are usually necessary to justify rigorously the manipulations customary
in treating problems of radiation.}
and smoothness of the worldline (at least $C^3$). 

Despite the great confusion and controversy 
in the literature over the presumed  
behavior of uniformly accelerated charged particles,
all authors seem to agree that a particle of charge $q$ accelerated 
(not necessarily uniformly) for a finite time interval
$[\tau_i, \tau_f]$ in Minkowski space
does radiate energy, given quantitatively by
\begin{equation}
\label{eqA4}
\mbox{energy radiation} = \frac{2}{3} q^2 \int_{\tau_i}^{\tau_f}  
- \atilde^2 \, d\taut 
\q.
\end{equation}
Note that the integrand is positive whenever $\atilde \neq 0$,
and the radiation can be made arbitrarily large
for a given uniform acceleration by making $\tau_f - \tau_i$
sufficiently large. 

The Lorentz-Dirac equation is equation \re{hobbseq} for Minkowski space
(implying zero Ricci tensor). 
It states that the four-force on the particle is the external Lorentz force
$q{F^i}_\alpha u^\alpha$ plus the  
Lorentz-Dirac radiation reaction term
 \begin{equation}
\label{ldrad}
 \frac{2}{3}q^2 
\left[ 
\frac{d\tilde{a}^i}{d\tilde{\tau}} + \tilde{a}^2 \tilde{u}^i 
\right]
\q.
\end{equation}
But \re{eqA3} states that this radiation reaction term vanishes
outside the transition intervals $[\taut_i , \taut_i + \epsilon]$
and $[\taut_f - \epsilon, \taut_f]$.
Hence those who believe that the Lorentz-Dirac term \re{ldrad} 
correctly describes the radiation reaction must admit the strange 
consequence that {\em the radiation reaction 
occurs {\bf only } in the 
transition intervals} at the beginning and ending of the trip. 
No matter how long the trip (i.e., no matter how large $\taut_f - \taut_i$),
the radiation reaction is confined to two short time intervals
of length $\epsilon$. 

Note that radiation reaction is not a hypothetical quantity;
in principle, 
it can be physically measured as the rate of fuel consumption of the rocket.
Thus belief in the correctness of the Lorentz-Dirac radiation reaction
requires the belief that the pilot of a charged rocket uniformly 
accelerated for a finite time observes {\em no fuel consumption
during the uniform acceleration}. All the radiated energy is paid for 
by fuel consumed in the beginning and ending transition intervals.

Since the energy radiated during 
the beginning transition interval $[\taut_i, \taut_i + \epsilon]$ 
is presumably independent of the length of the uniform acceleration,
for a very long uniform acceleration, most of the radiated energy 
is ``borrowed'', to be inexorably paid for at the end of the trip. 

What if the rocket doesn't carry enough fuel to pay for the borrowed radiated
energy (which can be made arbitrarily large by extending the period
of uniform acceleration)? 
This problem is solved in detail in  \cite{parrott2},
and it turns out the rocket mass goes negative.  
The borrowed energy is paid for by negative mass at the end of the trip!

If we disallow negative mass as unphysical, then the Lorentz-Dirac equation of 
motion implies that the trip is possible only if the initial mass (fuel)
is sufficient to pay for the radiated energy.
After a small down payment for the initial transition period
from inertial to uniformly accelerated motion,
the radiated energy is ``borrowed'' until the end of the trip,
at which time the debt is paid. 
If the initial mass is insufficient to pay the debt, 
the trip is presumably impossible,
but under the assumption that the Lorentz-Dirac radiation reaction
term identifies with the rate of fuel consumption,
we don't find this out until the end of the trip!

Since this sounds so unlikely, 
it should be emphasized that it is a mathematically rigorous
consequence of identification of the energy radiation rate 
given by the Lorentz-Dirac radiation reaction term
with fuel consumption.
There are no approximations whatever in the argument leading to it. 
To my knowledge, it has never been questioned.% 
\footnote{Indeed, the paper http://gr-qc/9303025 was rejected by several
journals on the grounds that it is too trivial.
No substantive objections have been raised to its mathematics. 
Nor was it considered one of those papers too vague to be judged
correct or incorrect;
several referees praised it as clearly written,
though not sufficiently novel or mathematically complicated
for their journals.}

There is no mathematical contradiction,
but it is hard to believe that this behavior would be seen 
in nature. This is one of several reasons that many are  skeptical
about the Lorentz-Dirac equation (along with its generalization,
the DBH equation).

We have the following situation:
\begin{enumerate}
\item
To the best of my knowledge,
all authors agree that a charged particle uniformly accelerated for 
a finite time in Minkowski space does radiate in accordance with \re{eqA4}.
\item
Some authors (e.g., Singal \cite{singal})
believe that nevertheless, a charged particle in Minkowski space
uniformly accelerated for {\em all} time would {\em not} radiate.
Indeed, Quinn and Wald \cite{quinn/wald97} 
take this as an {\em axiom}, from which 
they obtain the DBH equation.
\item
My opinion is that the question of whether 
radiation would be observed from
a charged particle
uniformly accelerated for all time is, according to taste,
either meaningless or a matter of arbitrary definition. 
The reasons are given briefly in \cite{parrott3}, and more fully
in \cite{parrott2}.  See also \cite{parrott}, Chapters 4 and 5, for background. 
\end{enumerate}

The identification of geodesic motion in the $g$-spacetime
with uniform acceleration in Minkowski space, and the fact that the
corresponding radiation reaction terms vanish in both contexts,
makes it unsurprising that there should exist in $g$-spacetime analogs  
of the crazy consequences of the Lorentz-Dirac equation for uniform
acceleration in Minkowski space.  
The example of Section 4 is one such analog.  
There are other analogs which fit more smoothly into the framework
of the energy conservation analysis of Quinn and Wald \cite{quinn/wald},
but these require too much detailed knowledge of the Quinn/Wald setup
to be worth presenting here. 

However, it is not clear that there is any detailed correspondence
between a charged particle uniformly accelerated for a finite time
and satisfying the Lorentz-Dirac equation in Minkowski space and  
a particle undergoing geodesic motion for a finite time and satisfying
the DBH equation in the $g$-spacetime of Section 4.
The reason is that although the DBH radiation reaction term 
\re{eqA1} is conformally
invariant, the DBH equation \re{hobbseq} itself is not 
(because the left side is not conformally invariant). 
Sonego \cite{sonego} discusses noninvariance of the DBH equation in more detail.

Finally, I want to note some deficiencies in the present work.  
It implicitly assumes that the fields in a conformally flat spacetime
are the same as in Minkowski space, and this assumption should have
been stated explicitly.  
Due to the conformal invariance of the distributional
Maxwell equations,
the Minkowski space fields do satisfy Maxwell's equations in any conformally
flat spacetime, but the uniqueness of such solutions seems not to have
been rigorously established, though intuitively it is expected.  
Other authors (e.g., Hobbs \cite{hobbs}), appear to make the same assumption,
so perhaps it can be justified in some way unknown to me. 

The argument on pages 5 and 6 leading to the conclusion
that $\atilde^2 $ cannot vanish identically (for a freely falling particle 
starting at rest in the described conformally flat space time)
is correct as stated, but its conclusion is not as strong as needed 
for the rest of the paper.
The problem is that in order to unambiguously identify  
the energy radiation as given by \re{eq3},
one needs to assume that the particle was unaccelerated 
in distant past and future, and that its worldline is smooth.% 
\footnote{I assume $C^\infty$ for mathematical simplicity, 
and this should have been explicitly stated in the original.
With carefully chosen definitions, $C^2$ would be enough.} 
If the particle is at rest up to the time the free fall starts,
then its worldline fails to be differentiable at the starting time.  
This deficiency can be repaired by inserting a smoothing transition interval
and invoking Sonego's result \cite{sonego} 
to conclude that in fact, 
$\atilde^2$
is a nonzero {\em constant} during the free fall. 
\subsection{Discussion of the proofs of \cite{sonego} and \cite{quinn/wald}} 

The literature of relativistic electrodynamics is notoriously unreliable.
The problems are physically subtle, and the mathematics tends to be
complicated, with much tedious algebra. 
Errors in the literature are common 
and rarely corrected. 
I have found that the only way to be sure of a result is to 
check it oneself in detail, including the tedious algebra. 

The main purpose of this appendix is to put in proper context
the plausibility argument of the original. 
Although the original analysis still seems basically valid,
the subsequent work of Sonego shows that 
it did not tell the whole story. 
A secondary purpose is to share with those interested in such 
problems my opinions concerning the mathematics of  
the proofs of Sonego and of Quinn and Wald.  

The DBH radiation reaction term \re{eqA1} applies only to conformally flat
spacetimes.  For general spacetimes, there is an additional term
called the ``tail term''.  Sonego calls the term \re{eqA1}
the ``local term''.  The full DBH radiation reaction is the sum 
of the local term and the tail term.  
Sonego concludes that the local term and the tail term are separately 
conformally invariant.   

I have checked his proof of conformal invariance of the local term.%
\footnote{Despite its appearance, this is not routine.
One indication is the fact that it has escaped notice for 
the thirty years since Hobbs corrected 
the original DeWitt/Brehme equation.} 
I regard this as a rigorously proved theorem.

I would hesitate to characterize conformal invariance of the tail
term as a mathematically rigorous theorem.
Sonego's argument seems to require auxiliary assumptions for which I 
haven't been able to find proofs.
However, these assumptions are plausible, and I would expect
some version of conformal invariance of the tail term 
to be rigorously provable,
possibly under auxiliary technical hypotheses.

The present work (a copy of which was sent to Quinn and Wald
in 1996) raised the question of whether the DBH equation 
conserves ``energy'' as defined by the usual construction 
 \re{eq2.5} 
in spacetimes with a timelike Killing vector.  
The 1999 paper of Quinn and Wald \cite{quinn/wald} 
answers this question by
presenting a proof 
that indeed it does conserve energy.
This is not in contradiction to the example of Section 4
for reasons explained above.  
Even if the DBH equation of motion equation conserves energy, 
for the geodesic motion
for a finite time considered in Section 4, 
all the radiated energy is furnished by radiation reaction 
at the beginning and ending of the trip.
For example, in the trip from $A$ to $B$ of Figure 3, 
all the radiated energy is furnished in the transition regions
near $A$ and $B$ where the Ricci tensor does not vanish.%
\footnote{The original example should have included a transition
region near $A$ in order to make the worldline smooth as 
mentioned above.  Also, the curve depicted in Figure 3 
should have started at $A$ (rather than being extended 
to the left of $A$) and ended at $C$.  
} 

There is a major gap in the Quinn/Wald proof around their equation (42).
In private correspondence, the authors have convinced me that
it can probably be filled. 
However, the details of the repair are likely to be complicated,
and I do not know if they have been written out. 

It looks to me as if their equations (39) and (18) may be in error.
If so, the errors are potentially serious enough to invalidate
the paper's main result.  
The authors have not answered a letter of October, 1999, 
enquiring how to justify these equations,
nor was a followup letter of December (1999) answered.
(This is being written in July, 2000.)

	So, I cannot vouch for the correctness of the Quinn/Wald proof.
However, after a careful study of much of the paper,
I can vouch for its overall interest. 
It is clearly written and contains many potentially useful new ideas.
I have learned much from it, and 
it will probably repay study for 
anyone seriously interested in fundamental problems
of electrodynamics in curved spacetimes. 
\subsection{Reexamination of the original conclusions}
The body of the paper is essentially the same as that posted in 1993.  
It was rejected by several journals on the grounds that it is mathematically 
too trivial,
a judgment which I am not in a position to dispute.
The only referee who saw any problem with its mathematics or conclusions
was one who questioned whether 
Maxwell's equations were conformally invariant!  
(That was his only objection.)
 
However, in the light of Sonego's result, it is clear that 
its focus was to some degree misdirected.  
I found striking the fact that although radiation in a conformally flat
spacetime could be computed as if it were Minkowski space,
the DBH radiation term appeared to depend on the conformal factor.
At the time, given known deficiencies in derivations
of the DBH equation, this seemed presumptive evidence  
that ``the DeWitt/Brehme/Hobbs equation should not be
uncritically accepted'', evidence which seemed to go beyond
the usual objections to the Lorentz-Dirac radiation reaction term. 
Sonego's proof that the DBH radiation reaction term 
is conformally invariant 
shows that this presumptive evidence was only illusory.%
\footnote{In retrospect, it seems that an opportunity to conjecture 
Sonego's result was missed here.  
Whatever the other questionable aspects of the DBH derivation,
it was carried out in a covariant way starting from the conformally
invariant Maxwell equations, so it would seem reasonable 
to conjecture that the result would be conformally invariant.}
(However, other objections to the DBH equation are unaffected.)

\end{document}